\documentclass[aps,pra,twocolumn,10pt,superscriptaddress,amsmath,amssymb]{revtex4-2}  % Include the cls file since revtex 4.2 is not supported by arxiv

\usepackage[T1]{fontenc}
\usepackage[utf8]{inputenc}
\usepackage[a-2u,mathxmp]{pdfx}  % Try to ensure PDF/A standards
\usepackage[pdfa]{hyperref}  % Hyperlinks within the pdf
\usepackage[stretch=10,shrink=10,final]{microtype}  % Nicer microtypography
\usepackage{graphicx}  % Include figure files
\usepackage{bm}  % Bold math
\usepackage{acro}  % Abbreviations, include the sty file for acro 3.3 since the version on arxiv has a bug using the "single" setting
\usepackage{subcaption}  % Subfigures
\newcommand{\figref}[1]{Fig.~\ref{fig:#1}}
\newcommand{\secref}[1]{Sec.~\ref{sec:#1}}

\newcommand{\refref}[1]{Ref.~\cite{#1}}
\renewcommand{\eqref}[1]{Eq.~(\ref{eq:#1})}
\renewcommand{\tablename}{Table}
\renewcommand{\figurename}{Figure}

\newcommand{\ket}[1]{\left|{#1}\right\rangle}

\newcommand{\braket}[2]{\left \langle #1 \middle| #2 \right \rangle}
\newcommand{\braketmatrix}[3]{\left \langle #1 \middle| #2 \middle| #3 \right \rangle}

\newcommand{\tsp}{TS\raisebox{0.2ex}{+}1}

\newcommand{\chillijl}{\textsc{chilli.jl}}
\newcommand{\julia}{\textsc{Julia}}
\newcommand{\libxc}{\textsc{libxc}}
\newcommand{\pyscf}{\textsc{PySCF}}
\newcommand{\pyflosic}{\textsc{PyFLOSIC2}}
\newcommand{\pysis}{\textsc{pysisyphus}}
\newcommand{\pycom}{\textsc{PyCOM}}
\newcommand{\obabel}{\textsc{Open Babel}}

\acsetup{single,load-style=abbrev}

\begin{document}
\title{Bond formation insights into the Diels-Alder reaction:\texorpdfstring{\\}{}A bond perception and self-interaction perspective}
\author{Wanja Timm Schulze}
\email{wanja.schulze@uni-jena.de}
\affiliation{Institute for Physical Chemistry, Friedrich Schiller University, Jena, 07743, Germany}
\author{Sebastian Schwalbe}
\affiliation{Institute of Theoretical Physics, TU Bergakademie Freiberg, Freiberg, 09599, Germany}
\affiliation{Center for Advanced Systems Understanding (CASUS), G\"orlitz, 02826, Germany}
\affiliation{Helmholtz-Zentrum Dresden-Rossendorf (HZDR), Dresden, 01328, Germany}
\author{Kai Trepte}
\affiliation{Taiwan Semiconductor Manufacturing Company North America, San Jose, USA}
\author{Alexander Croy}
\affiliation{Institute for Physical Chemistry, Friedrich Schiller University, Jena, 07743, Germany}
\author{Jens Kortus}
\affiliation{Institute of Theoretical Physics, TU Bergakademie Freiberg, Freiberg, 09599, Germany}
\author{Stefanie Gräfe}
\affiliation{Institute for Physical Chemistry, Friedrich Schiller University, Jena, 07743, Germany}

\date{\today}

\begin{abstract}
\noindent
The behavior of electrons during bond formation and breaking cannot commonly be accessed from experiments.
Thus, bond perception is often based on chemical intuition or rule-based algorithms.
Utilizing computational chemistry methods, we present intrinsic bond descriptors for the Diels-Alder reaction, allowing for an automatic bond perception. We show that these bond descriptors are available from localized orbitals and self-interaction correction calculations, e.g., from Fermi-orbital descriptors. The proposed descriptors allow a sparse, simple, and educational inspection of the Diels-Alder reaction from an electronic perspective. We demonstrate that bond descriptors deliver a simple visual representation of the concerted bond formation and bond breaking, which agrees with Lewis' theory of bonding.
\end{abstract}

\acresetall

\maketitle

\section{Introduction \label{sec:intro}}
Bond type perception, the identification of bond types in a given molecule, is a longstanding problem in many modern-day applications, including visualization \cite{Lazzari2020}, force field calculations \cite{Zhang2012}, or multiscale methods \cite{Seeber2022}. While a human interpretation of bonding situations appears trivial in many cases, an \textit{ab initio} description of bonding based on computational chemistry can be more challenging. Over the years, various simple descriptions of bonding have been developed, like Lewis' theory \cite{Lewis1916} or Linnett's double-quartet theory \cite{Linnett1960, Linnett1961, Luder1966, Empedocles1964}. Alternatively, heuristic criteria are often used for bond perception, e.g., the atomic distance being smaller than the sum of covalent radii. Such criteria can be used to determine bonded atoms while additional rule-based algorithms can be employed to detect bonds to next-nearest neighbors in the local environment of an atom \cite{Wang2006, Zhang2012, Artemova2016}.

Depending on the application, such criteria may be too simple to accurately describe the bond order for any molecule of interest. More sophisticated methods are for example Bader's \ac{aim} analysis \cite{Bader1991, Bader1994}, where the Laplacian of the electronic density is analyzed, or the \ac{elf} \cite{Becke1990} that also accounts for the Pauli exclusion principle. Regardless, bond orders cannot directly be obtained from either method. Other methods provide bond order analysis, e.g., Wiberg- \cite{Wiberg1968} or Mayer-bond orders \cite{Mayer2007}. Nonetheless, one often obtains fractional bond orders making a classification into single, double, or triple bonds cumbersome.

An alternative description of bond orders can be found with the use of so-called \acp{fod} \cite{Schwalbe2019}. In recent research, it has been discussed that the \acp{fod} carry chemical bonding information and can be interpreted according to Lewis' or Linnett's double-quartet theory \cite{Schwalbe2019, Trepte2021}. Further, it has been shown that the center of mass (centroids) of different localized orbitals can be used to get a reasonable initial guess for the \ac{fod} positions \cite{Schwalbe2019} and to track bond rearrangements, e.g., by creating \textit{curly arrows} as a representation of bond reorganization during a reaction \cite{Klein2019, Sciortino2019}.

\begin{figure}[tbp]
	\centering
	\includegraphics[width=0.9\linewidth]{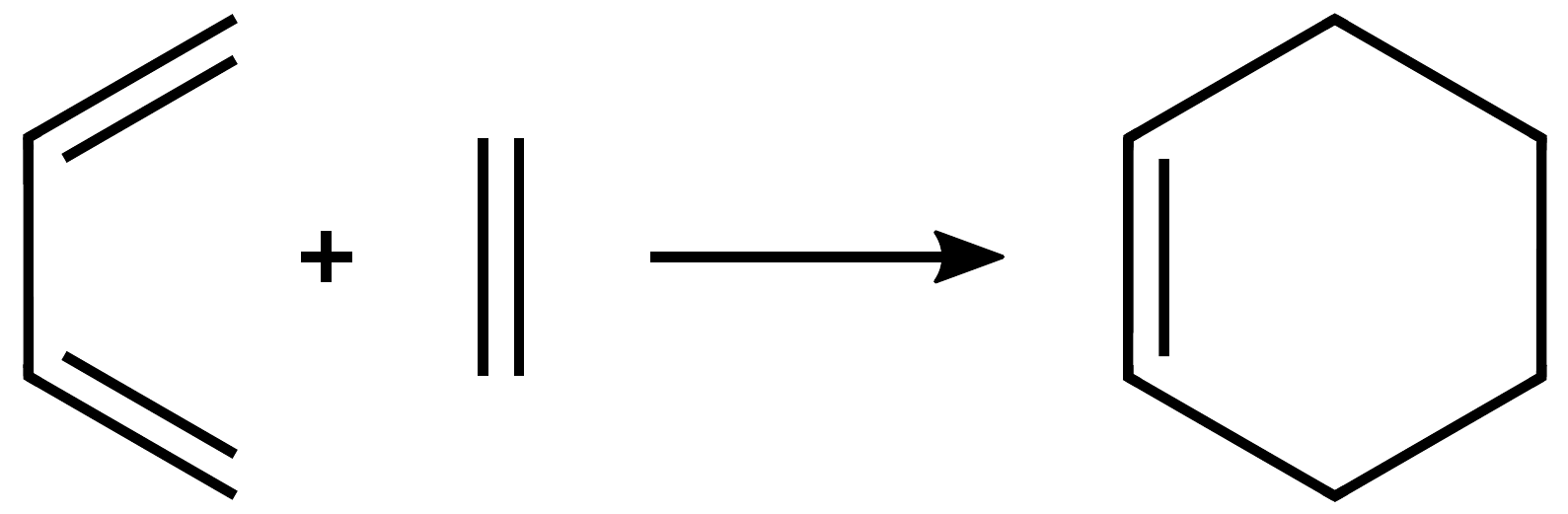}
	\caption{Schematic Diels-Alder reaction of 1,3-butadiene (C$_4$H$_6$) and ethylene (C$_2$H$_4$) to cyclohexene (C$_6$H$_{10}$). Two new carbon-carbon bonds will be created by breaking two double bonds and relocating another one. All bond changes happen simultaneously due to the concerted nature of the reaction.}
	\label{fig:reaction}
\end{figure}

Sometimes it may be desirable to track the bonding situation or even the bond formation and breaking of molecules in a reaction. This manuscript presents an analysis of bond descriptors for a chemical reaction, i.e., bond formation and bond breaking along a reaction path using automatic electron perception methods. As an example, we will model a well-known reaction -- the Diels-Alder reaction \cite{Diels1928, Alder1942, Hershberg1937}. Despite often being used as a textbook example for a concerted reaction mechanism where all bond changes happen simultaneously \cite{Townshend1976, Houk1986, Bernardi1988, Goldstein1996}, this reaction is still very important in modern applications, such as self-healing materials \cite{Geitner2015, Ratwani2023}. In the simplest form of this reaction, 1,3-butadiene reacts with ethylene to form cyclohexene, as displayed in \figref{reaction}. In this work, we will provide insights into the bond formation for the Diels-Alder reaction from a bond perception and self-interaction perspective.

This manuscript is structured as follows: In \secref{theory} the theoretical background is introduced. The computational details and methodology to model the Diels-Alder reaction are described in \secref{comp}. The results are presented in \secref{res}. The calculated \acp{fod} and centroids will be shown and compared. Afterwards, energies and other properties that have been calculated along the reaction path will be presented. The corresponding information on bond formation and breaking is discussed. Subsequently, the validity of the qualitative results is verified. The results are summarized in \secref{summary}.
\clearpage

\section{Theoretical background \label{sec:theory}}
Given its moderate computational effort and sufficient accuracy, \ac{dft} \cite{Hohenberg1964, Kohn1965} has become one of the most commonly employed methods in computational chemistry and related research fields \cite{Mourik2014}. The total energy of a system is given by the \ac{ks} energy functional
\begin{align}
\begin{split}
	E_{\mathrm{KS}}[n^\alpha,n^\beta] &= T_\mathrm{s}[n^\alpha,n^\beta] + V_\mathrm{ext}[n] \\
    &\quad + E_\mathrm{H}[n] + E_\mathrm{XC}[n^\alpha,n^\beta],
\end{split}
\end{align}
with $T_\mathrm{s}$ as the kinetic energy, $V_\mathrm{ext}$ as the external potential, $E_\mathrm{H}$ as the Coulomb energy, and $E_\mathrm{XC}$ as the exchange-correlation energy. In an open-shell description of a molecular system and related unrestricted calculations, there are two spin channels $\sigma$, i.e., the $\alpha$ and $\beta$ spin channels. Accordingly, $n$ denotes the total electronic density while $n^\sigma$ denotes the spin density.

While being formally exact, for practical \ac{dft} calculations additional assumptions regarding the exchange-correlation functional are needed. With those assumptions and further numerical approximations, reasonable results for a large variety of systems can be obtained \cite{Burke2012}. However, those \acp{dfa} can also introduce unwanted side effects, such as the artificial interaction of electrons with themselves \cite{Perdew1981}.
Since for one-electron systems the Coulomb energy ($E_\mathrm{H}$) and exchange-correlation contributions ($E_\mathrm{XC}$) have to cancel each other out, violating this condition will result in the so-called \ac{sie} $E_\mathrm{SI}$
\begin{align}
	E_\mathrm{SI}[n^\sigma_1] = E_\mathrm{H}[n^\sigma_1] + E_\mathrm{XC}[n^\sigma_1, 0],
\end{align}
where $n^\sigma_1$ is a single-electron density for an electron with spin $\sigma$.
Recent research \cite{Schwalbe2022} showed that the \ac{oee} \cite{Lonsdale2020, Lonsdale2022}, i.e., an error closely related to \ac{sie}, is still dominant in modern exchange-correlation functionals.

In the formulation of \ac{pz} \cite{Perdew1981}, the \ac{sie} is removed from the \ac{ks} energy $E_\mathrm{KS}$ for all $N^\sigma$ occupied orbitals, resulting in a \ac{sic}
\begin{align}
	E_\mathrm{PZ}[n^\alpha,n^\beta] = E_\mathrm{KS}[n^\alpha,n^\beta] - \sum_\sigma\sum_i^{N^\sigma} E_\mathrm{SI}[n^\sigma_i].
	\label{eq:pzsic}
\end{align}
The \ac{flosic} method \cite{Pederson2014, Pederson2015, Pederson2015a, Yang2017} is a novel form of \ac{pzsic}, where \acp{flo} are used to calculate the one-electron densities. \acp{fo} can be obtained by applying the transformation matrix $\bm R^\sigma$ to the molecular orbital coefficients $\bm C^\sigma$
\begin{align}
	\bm c^{\mathrm{FO},\sigma} = \bm C^ \sigma \bm R^\sigma.
\end{align}
The transformation matrix $\bm R^\sigma$ can be obtained with
\begin{align}
	R_{ij}^{\mathrm{FO},\sigma} = \frac{
		\braket{\psi_j^{\sigma}}{\bm a_i^{\sigma}}}{\sqrt{n^{\sigma}(\bm a_i^{\sigma})}}.
\end{align}
Here, the $\psi_i$ are occupied orbitals, while $\ket{\bm a_i^{\sigma}}$ denote position eigenstates localized at the \acp{fod} $\bm a_i^{\sigma}$, which are reference electron positions to build the \acp{fo}. In \ac{flosic} calculations these \acp{fod} are optimized along with the electronic density. The \acp{flo} are obtained by applying Löwdin's symmetrical orthonormalization \cite{Loewdin1950} to the \acp{fo}.
The resulting \acp{flo} are well-localized, comparable to other localized orbitals, e.g., \ac{fb} orbitals \cite{Foster1960} where the orbital variances $\mathcal{J}_\mathrm{FB}$ are minimized \cite{Kleier1974}
\begin{align}
	\mathcal{J}_\mathrm{FB} = \sum_i \left( \braketmatrix{\psi_i}{\bm r^2}{\psi_i} - \braketmatrix{\psi_i}{\bm r}{\psi_i}^2 \right).
	\label{eq:spread}
\end{align}
The lower the value of $\mathcal{J}_\mathrm{FB}$ the more localized is the state.
It is known that \ac{sic} raises reaction energy barriers \cite{Johnson1994, Patchkovskii2002, Johansson2008}. For localized \ac{sic} orbitals this can be explained by the increased noding of valence orbitals in the \ac{ts} that lowers the \ac{sie} (or raises the correction) \cite{Shahi2019}.
The novel aspect of our contribution is the explicit analysis of how \ac{sic} and information derived from localized orbitals can be used to track the bond formation along a reaction path.

\section{Computational details \label{sec:comp}}
\subsection{Electronic structure codes}
For transparent results, only open-source codes were used in this manuscript following the \acf{foss} approach \cite{Lehtola2022}.
The all-electron \ac{gto} code \pyscf{} \cite{Sun2020} was used for all \ac{dft} calculations while the \pyflosic{} code \cite{Schwalbe2020, Liebing2022} was used for all \ac{flosic} calculations. The computational parameters for these calculations follow \refref{Trepte2021}. The local spin density
approximation Slater exchange functional \cite{Bloch1929, Dirac1930} with the modified Perdew-Wang correlation functional \cite{Perdew1992} (SPW92) has been used for all calculations, as implemented in the \libxc{} library \cite{Lehtola2018}. The double-$\zeta$ polarization consistent pc-1 basis set \cite{Jensen2001, Jensen2002, Jensen2002a} was employed. A \pyscf{} grid level of 7 was used. This corresponds to a (90,974) grid for hydrogen and a (135,1202) grid for carbon in the multi-center quadrature scheme \cite{Becke1988}. Pruning has been disabled for all calculations. The \ac{scf} energy convergence threshold was set to $1 \times 10^{-8}\,E_\mathrm{h}$. All calculations were performed spin-unrestricted.

To verify the results, \ac{dft} and \ac{flosic} calculations have been performed with a self-written \julia{} code called \chillijl{} \cite{Schwalbe2023}. A comparison of the calculated energies can be found in the supplementary material. The nuclear geometry optimizations, \ac{ts} optimization, and \ac{neb} calculations were carried out with the external optimizer \pysis{} \cite{Steinmetzer2021} using an interface to \pyscf{}. The computational parameters and methodology used will be explained in detail below.

\subsection{Nuclear geometry optimization}
In the target reaction, the reactant state contains the separated molecules ethylene and 1,3-butadiene. Along a reaction path, the reactant state transforms via a transition state into the product state, cyclohexene. Nuclear geometry optimizations were started from experimental geometries: The reactant ethylene was taken from \refref{Linstrom1997}. For butadiene, the gauche-1,3 geometry was taken from \refref{Baraban2018}. The structure for the product, cyclohexene, was taken from \refref{Chiang1969}. The initial structures were optimized at the \ac{dft} level of theory using the \ac{rfo} method \cite{Banerjee1985}.
The threshold for the maximum force $F_{\mathrm{max}}$ and the root-mean-square force $F_{\mathrm{rms}}$ have been set to $1.5 \times 10^{-5}\,E_{\mathrm{h}}/a_0$ and $1 \times 10^{-5}\,E_{\mathrm{h}}/a_0$, respectively.

Subsequently, the optimized nuclear geometries for ethylene and butadiene were combined, with the initial fragment distance being similar to \refref{Ramirez2015}. The resulting reactant state was optimized utilizing translational and rotational intrinsic coordinates \cite{Wang2016}, using the same force thresholds as before. It has been verified that the resulting Hessian of all nuclear geometry optimizations has only positive eigenvalues to ensure that stable minima have been found.

\subsection{Transition state optimization and reaction path}
To approximate the \ac{ts}, a nudged elastic band (NEB) calculation has been utilized at the \ac{dft} level of theory. The computational parameters were adapted from Refs. \cite{Herbol2017} and \cite{Ruttinger2022}. Eleven images have been calculated using the limited-memory BFGS (L-BFGS) optimizer \cite{Nocedal1980, Liu1989}, as implemented in \pysis{} \cite{Steinmetzer2021}. A maximum force of $2 \times 10^{-4}\,E_{\mathrm{h}}/a_0$ (${\approx}\,0.01\,$eV/\AA{}) and a step size of $0.075\,a_0$ (${\approx}\,0.04\,$\AA{}) were set. The climbing image was enabled, while the Kabsch algorithm \cite{Kabsch1976} to align images has been enabled. The splined \ac{hei} from the \ac{neb} calculation was used to optimize the \ac{ts}. This optimization uses the same thresholds as the previous nuclear geometry optimizations, using a restricted-step \ac{rfo} (\acs{rsirfo}) method \cite{Banerjee1985, Besalu1998}. After the \ac{ts} optimization, it has been checked that the resulting Hessian contains solely one imaginary frequency to ensure the result truly is a \ac{ts}.

In addition, two \ac{neb} calculations were performed with the same parameters as the previous \ac{neb} calculation, but with 9 images and the climbing image disabled. One \ac{neb} samples the \ac{mep} between the reactants and the \ac{ts}, and the other samples between the \ac{ts} and the product. This procedure resulted in 21 optimized geometries (1 reactant, 1 \ac{ts}, 1 product, and $2 \times 9$ \ac{neb} images).

\subsection{Orbital localization}
It is known that \acp{kso} tend to be delocalized over the whole extent of a given molecule \cite{Yang2017}. Thus, different localization schemes can be used to transform the \acp{kso} into localized orbitals. In contrast to \acp{kso}, localized orbitals can often be interpreted as orbitals carrying bond-related information \cite{Stewart2019}. The investigated localized orbitals are \ac{fb}, \ac{er} \cite{Edmiston1963}, \ac{pm} using L\"owdin charges \cite{Pipek1989}, generalized \ac{pm} using Becke charges \cite{Lehtola2014}, and \acp{ibo} \cite{Knizia2013}. All of these localization procedures were carried out using \pyscf{}, where the \ac{ciah} method is used \cite{Sun2017}, except for the \acp{ibo}, where the implementation and minimization follow \refref{Knizia2013} with an additional symmetric orthogonalization. As a simple stability analysis for the minimization with the \ac{ciah} method, the eigenvalues of the resulting Hessian have been calculated, and it has been ensured that all eigenvalues are positive.

\subsection{\ac{fod} optimization}
For the \ac{flosic} calculations, the initial \ac{fod} configuration were taken from the \pycom{} method \cite{Schwalbe2019}, i.e., the centroids of localized \ac{fb} orbitals. Note that the method was used with a stability analysis to ensure the local minima of the generated localized orbitals.
The initial \ac{fod} configurations follow Lewis' theory. With the need to optimize the density matrix and the \acp{fod}, the two-step \ac{scf} cycle following \refref{Karanovich2021} was employed as implemented in \pyflosic{}. The constrained L-BFGS-B method \cite{Byrd1995} was used for the \ac{fod} optimization. The final maximum force $F_{\mathrm{max}}$ acting on any \ac{fod} was below $1.5 \times 10^{-4} \,E_{\mathrm{h}}/a_0$.

\begin{figure*}[tbp]
    \begin{subfigure}{\textwidth}
		\centering
		\includegraphics[width=0.24\textwidth]{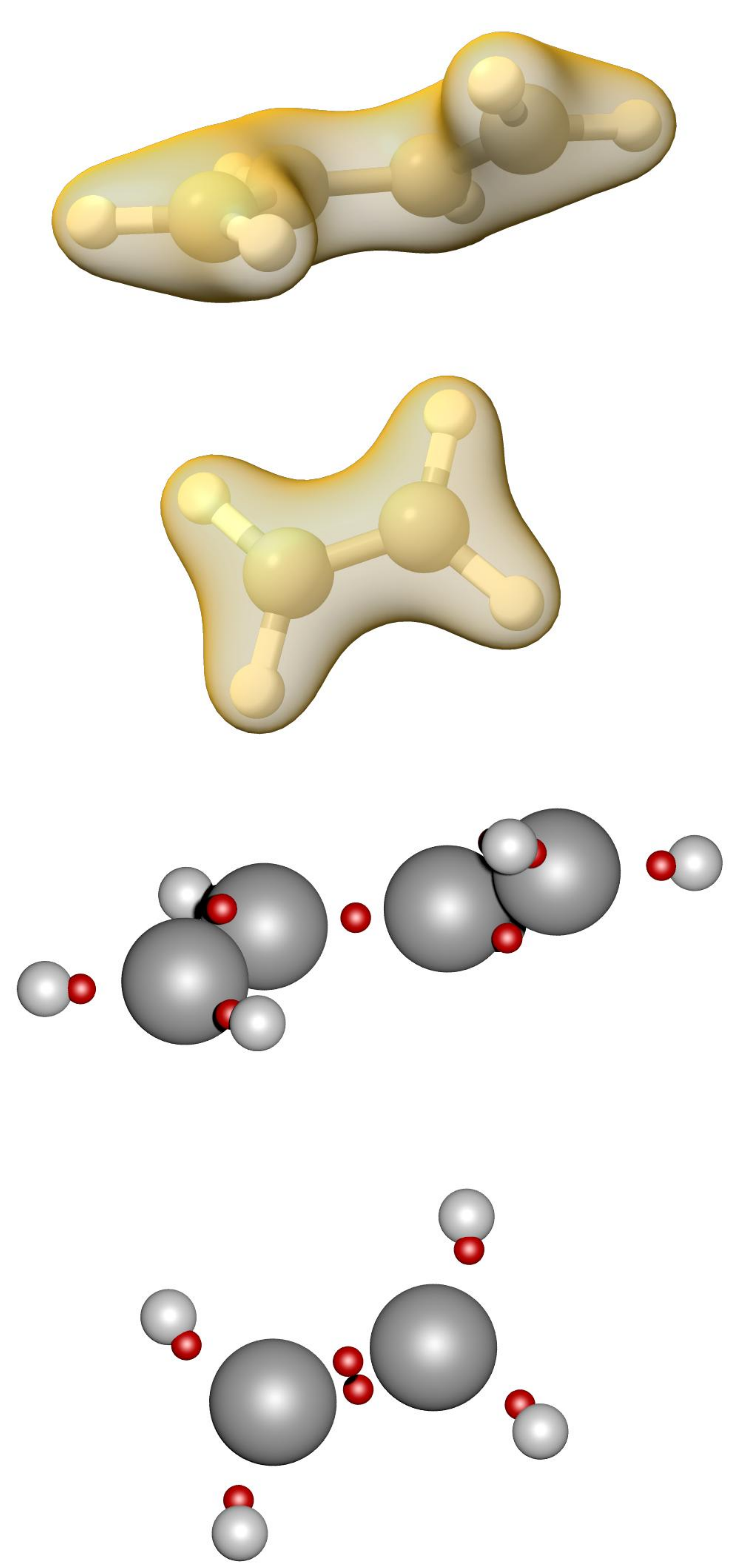}
		\includegraphics[width=0.24\textwidth]{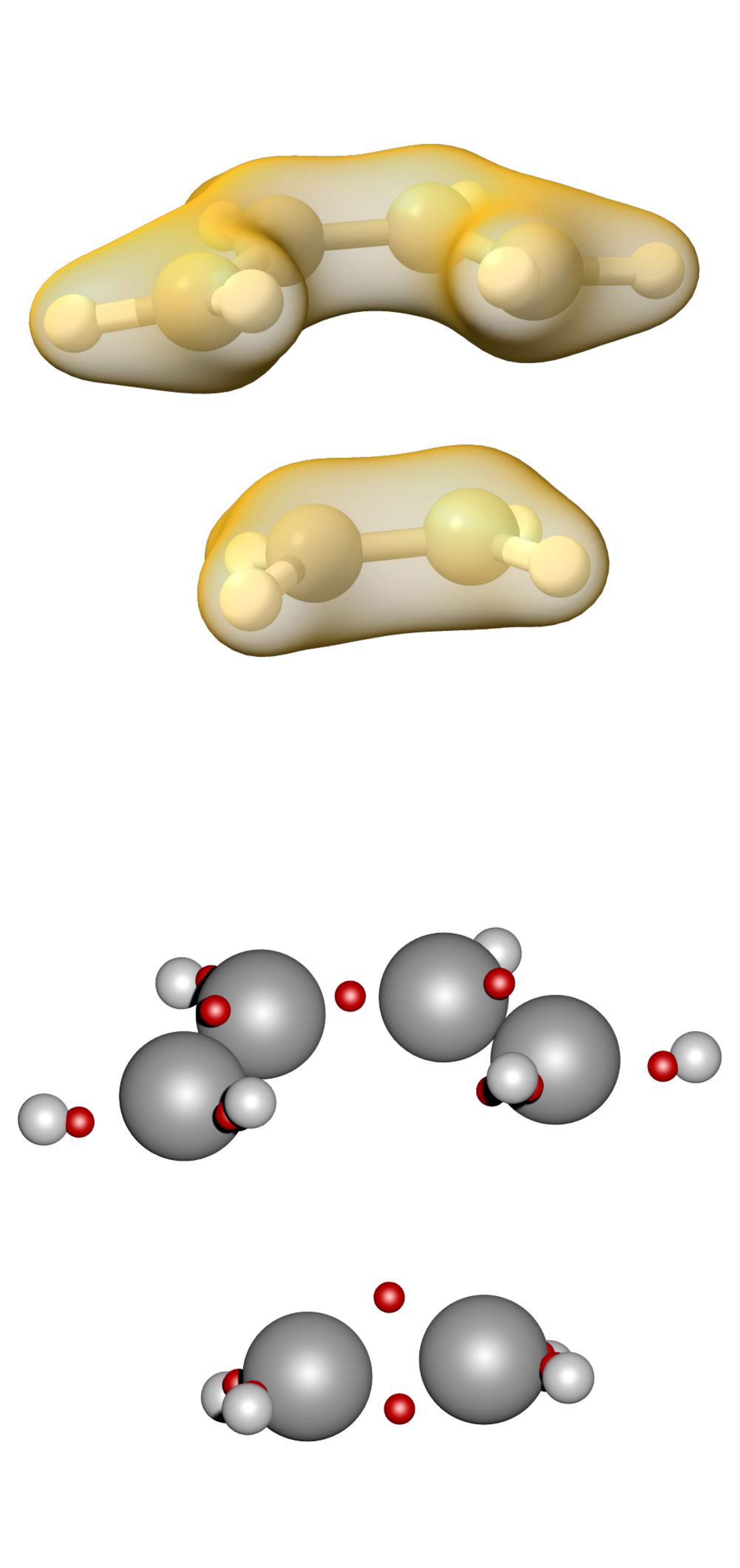}
		\includegraphics[width=0.24\textwidth]{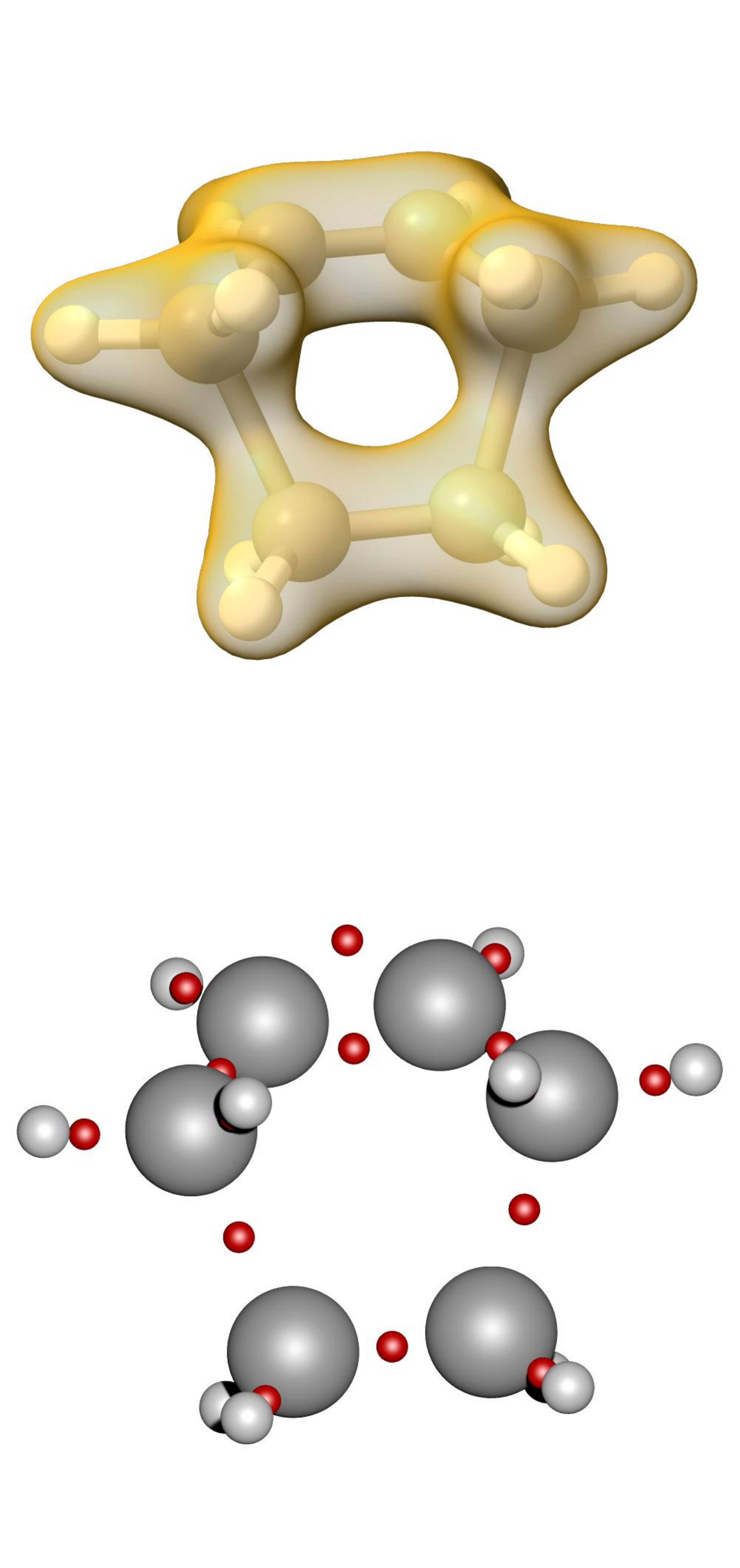}
		\includegraphics[width=0.24\textwidth]{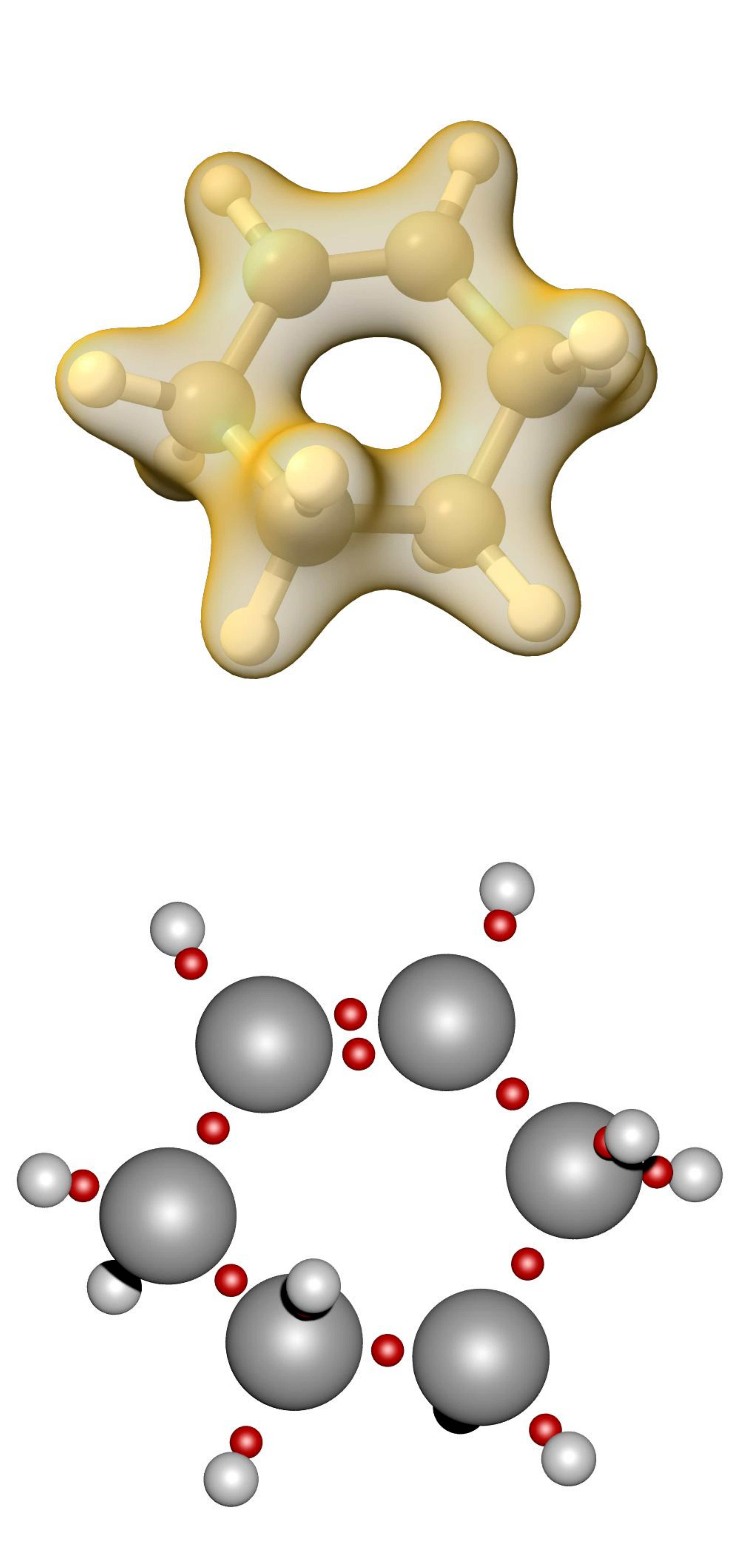}
		\subfloat[\label{fig:educt} Reactants]{\hspace{0.24\linewidth}}
		\subfloat[\label{fig:ts} \ac{ts}]{\hspace{0.24\linewidth}}
		\subfloat[\label{fig:image_11} \tsp{}]{\hspace{0.24\linewidth}}
		\subfloat[\label{fig:product} Product]{\hspace{0.24\linewidth}}
	\end{subfigure}

	\caption{Diels-Alder reaction of 1,3-butadiene (C$_4$H$_6$) and ethylene (C$_2$H$_4$) to cyclohexene (C$_6$H$_{10}$). Electronic density (top) and \acp{fod} (bottom) for the reactants, the \ac{ts}, the image after the \ac{ts} (\tsp{}), and the product. The isovalues have been chosen such that 85\,\% of the density is contained. More information about the selection of the isovalue can be found in the supplementary material. Carbon atoms are colored in grey, hydrogen in white, and the \acp{fod} in red. Only one spin channel has been displayed since the \ac{fod} positions for these images coincide for both spin channels. The centroids of localized orbitals resemble the \acp{fod}, see the supplementary material.}
	\label{fig:geometry}
\end{figure*}

\section{Results \label{sec:res}}
In this section, we present an overview of the main results, focusing on the bond analysis during the Diels-Alder reaction, employing spatial descriptors and other selected properties.

\subsection{Bond formation: Associated bond points}
This segment investigates the trends that can be found when analyzing the centroids of localized orbitals and \acp{fod}.
For simplicity, the centroids of localized orbitals and \acp{fod} positions will be labeled as \acp{abp}. As an example, the \acp{fod} have been displayed for selected structures in \figref{geometry}, along with the density. Corresponding figures using different centroids can be found in the supplementary material.

For the reactant (\figref{educt}) and product state (\figref{product}), the \acp{abp} correctly describe the anticipated bonding situation (compare with \figref{reaction}). Counting the \acp{abp} around each bond axis mimics the expected bond order \cite{Schwalbe2019} (see \figref{reaction}). Note that the number of \acp{abp} has to be divided by two since we are in the spin-unrestricted case. The \ac{ts} (\figref{ts}) still has the bond order of the reactant state. Over the course of the reaction, the bond order will change once, i.e., from the reactant state bond order to the product state bond order. This transition takes place one image after the \ac{ts} (\figref{image_11}) for all bonds and can be associated with the formation and breaking of bonds. This imagery of all bonds forming concertedly in one step is also in line with other research \cite{Townshend1976, Houk1986, Bernardi1988, Goldstein1996}.
We note that this behavior remains when evoking other localization methods, with small deviations. While the \acp{fod} and centroids of \ac{flo}, \ac{fb}, and \ac{er} orbitals are separated perpendicular to the bond axis, the centroids of \ac{pm} orbitals, generalized \ac{pm} orbitals and \acp{ibo} mostly lie directly on the bond axis.

As a measure of the change of the \acp{abp} during the reaction, one can use the distance $d_\mathrm{max}$, where $d_\mathrm{max}$ is the maximum distance of the \acp{abp} of bonding orbitals from the midpoint of the C-C bonds.
As the formation of two new C-C bonds, alongside the break up of two C-C double bonds over the course of the reaction necessitates a spatial rearrangement of \textit{localized} electrons pairs, this can be directly observed using $d_\mathrm{max}$.
A value of $d_\mathrm{max}\approx0$ would resemble a single bond, where the \ac{fod} positions would directly lie on the C-C bond axis. Typical distances for $d_\mathrm{max}$ for double bonds, i.e., in ethylene are about 0.4\,\AA{}, see the supplementary material for more details.

In \figref{fod_educt} one can see the $d_\mathrm{max}$ for the \acp{fod} for all breaking double bonds, starting from the initial reactant state towards the product structure. The \ac{fod} distance $d_\mathrm{max}$ remains almost constant in the beginning at 0.4\,\AA{}, as typical for double bonds. Approaching the \ac{ts} $d_\mathrm{max}$ increases, reaching its peak at the \ac{ts} (image 10). This means that \acp{fod} move farther away from the C-C bond axes. One can interpret the departing \acp{abp} as the breaking of a bond. After the \ac{ts}, the double bonds are broken into single bonds, with $d_\mathrm{max}$ being close to zero. This trend can be seen for all \acp{abp}. For the \ac{pm} orbitals and \acp{ibo}, one can see an oscillation of the distances for the centroids in ethylene. In addition, one can see that some orbitals behave differently depending on the spin channel, especially for \ac{pm} orbitals. Thus, in these cases, the symmetry is broken between $\alpha$ and $\beta$ spin channel. However, this behavior can not be found in the generalized \ac{pm} localization scheme. Since the trends between \ac{pm} and generalized \ac{pm} are otherwise the same, the oscillations and spin splitting are likely an artifact of the ill-defined L\"owdin charges \cite{Lehtola2014}.

As shown in the supplementary material, the values for $d_\mathrm{max}$ for the \ac{fb} centroids closely resemble the ones of the \acp{flo}. Thus, for the discussed reaction \pycom{} utilizing \ac{fb} orbitals is suggested to result in suitable initial \acp{fod}. The second best approximation for initial \acp{fod} utilizing \pycom{} would be achieved utilizing \ac{er}. \ac{pm}, generalized \ac{pm} and \acp{ibo} perform worse for the investigated reaction.

The formation of the new double bond in the product can be tracked as well, as seen in \figref{fod_product}. For the initial single bond, $d_\mathrm{max}$ is close to zero. The C-C single bond is present up to the \ac{ts}. The peak of $d_\mathrm{max}$ is at the \tsp{} where the new double bond has been formed. The distance stabilizes afterwards and reaches a value close to the previously observed 0.4\,\AA{} that is typical for a double bond.

\begin{figure}[bp]
	\begin{subfigure}{\linewidth}
		\centering
		\includegraphics[width=\linewidth]{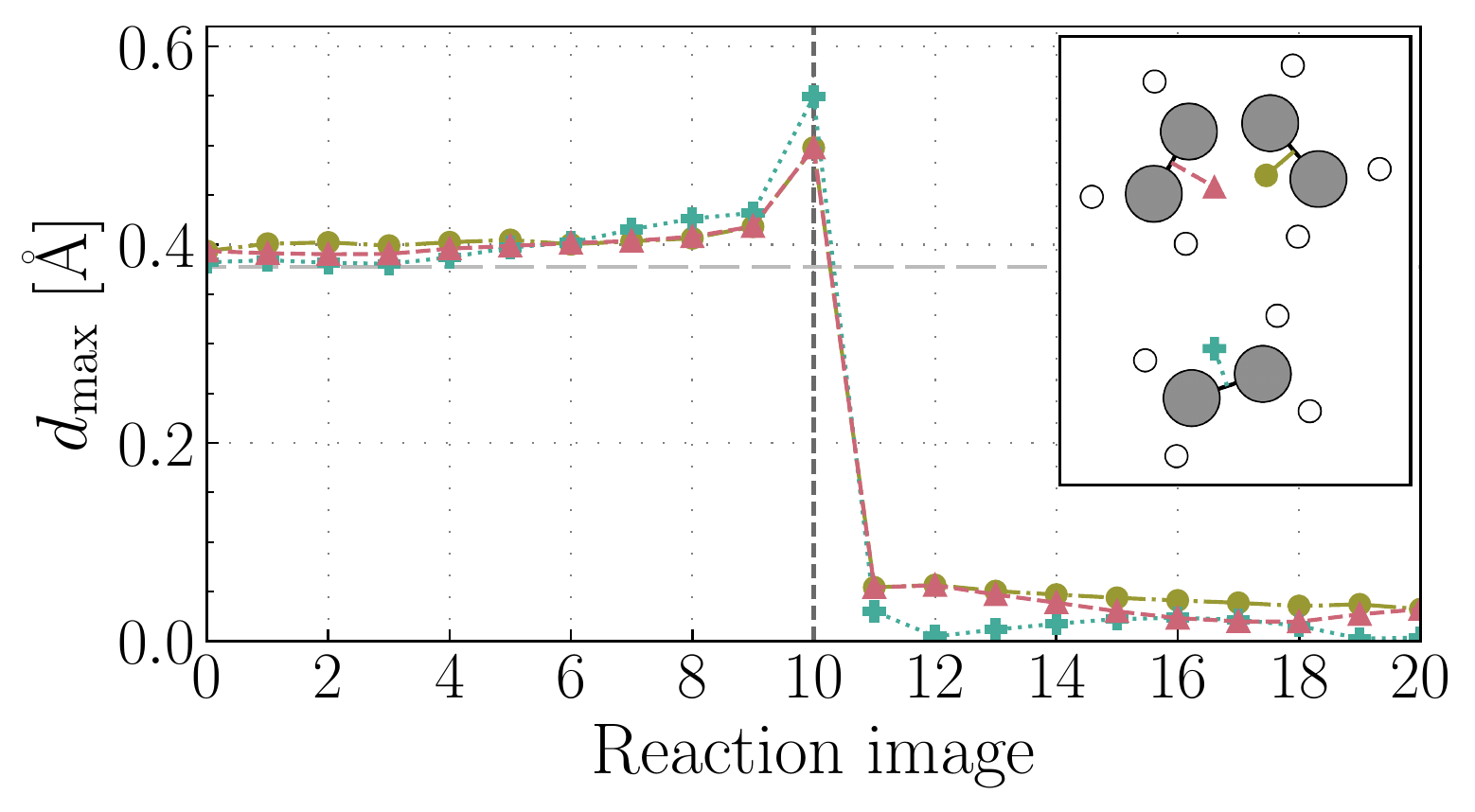}
		\caption{Maximum \ac{fod} distances for the breaking double bonds. Three different colors represent the selected \acp{fod}.}
		\label{fig:fod_educt}
	\end{subfigure}

	\begin{subfigure}{\linewidth}
		\centering
		\includegraphics[width=\linewidth]{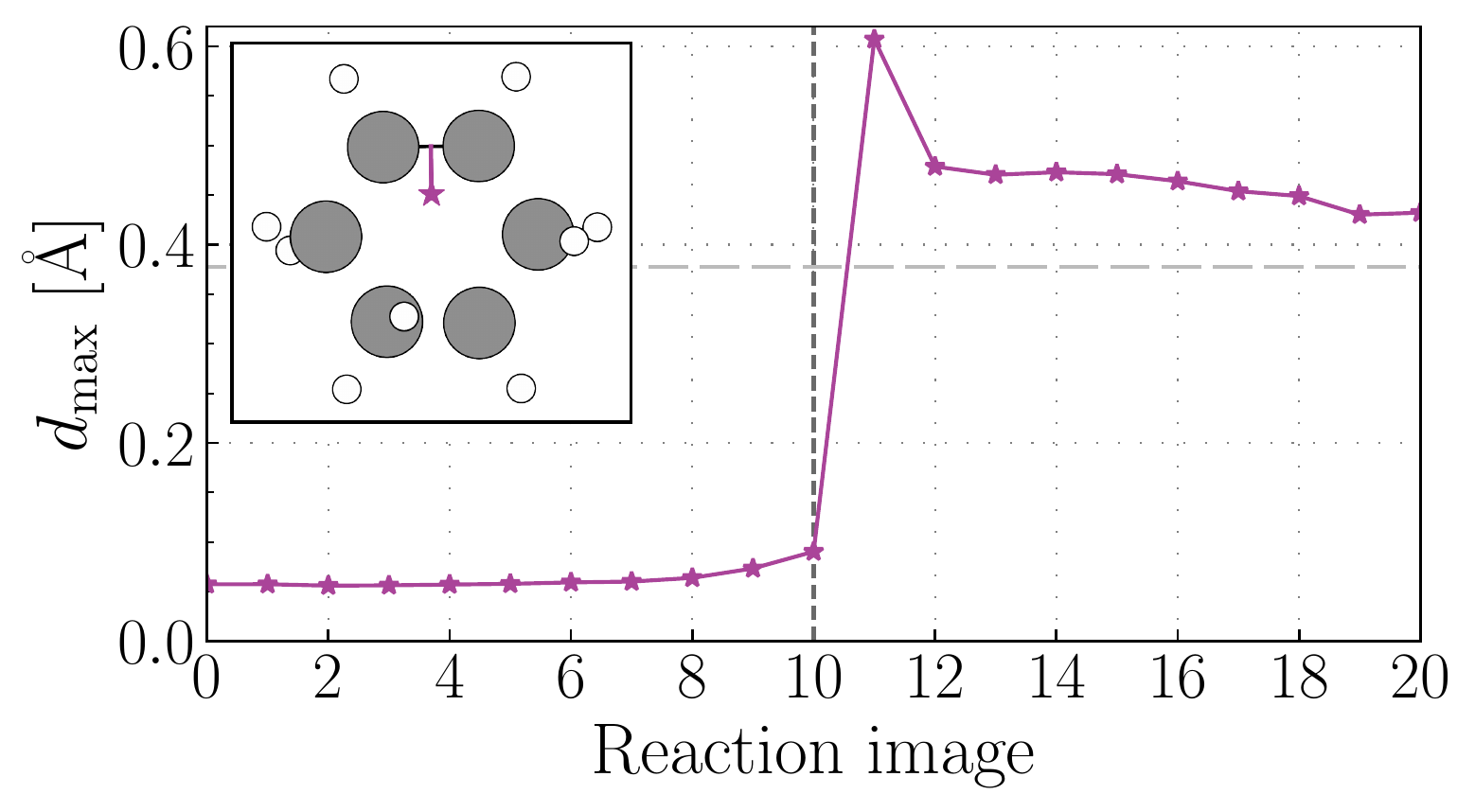}
		\caption{Maximum \ac{fod} distances for the forming double bond.}
		\label{fig:fod_product}
	\end{subfigure}

	\caption{Maximum \ac{fod} distances $d_\mathrm{max}$ from the C-C bond midpoints that form double bonds. The \ac{ts} is indicated at image 10 via a dashed vertical line. For comparison, the long dashed horizontal line indicates the \ac{fod} distance in ethylene. The color-coded \ac{fod} positions in the molecule are indicated schematically, where carbon atoms are colored in grey and hydrogen in white.}
	\label{fig:fod_educt_product}
\end{figure}

\subsection{Monitoring bond formation}
In this segment, scalar properties, e.g., total energies, \acp{sie}, and the absolute value of the dipole moment are analyzed along the calculated reaction path. We restrict ourselves to scalar properties, aiming to find simple scalar descriptors for bond formation and bond breaking.

\begin{figure}[hbtp]
	\begin{subfigure}{\linewidth}
		\centering
		\includegraphics[width=\linewidth]{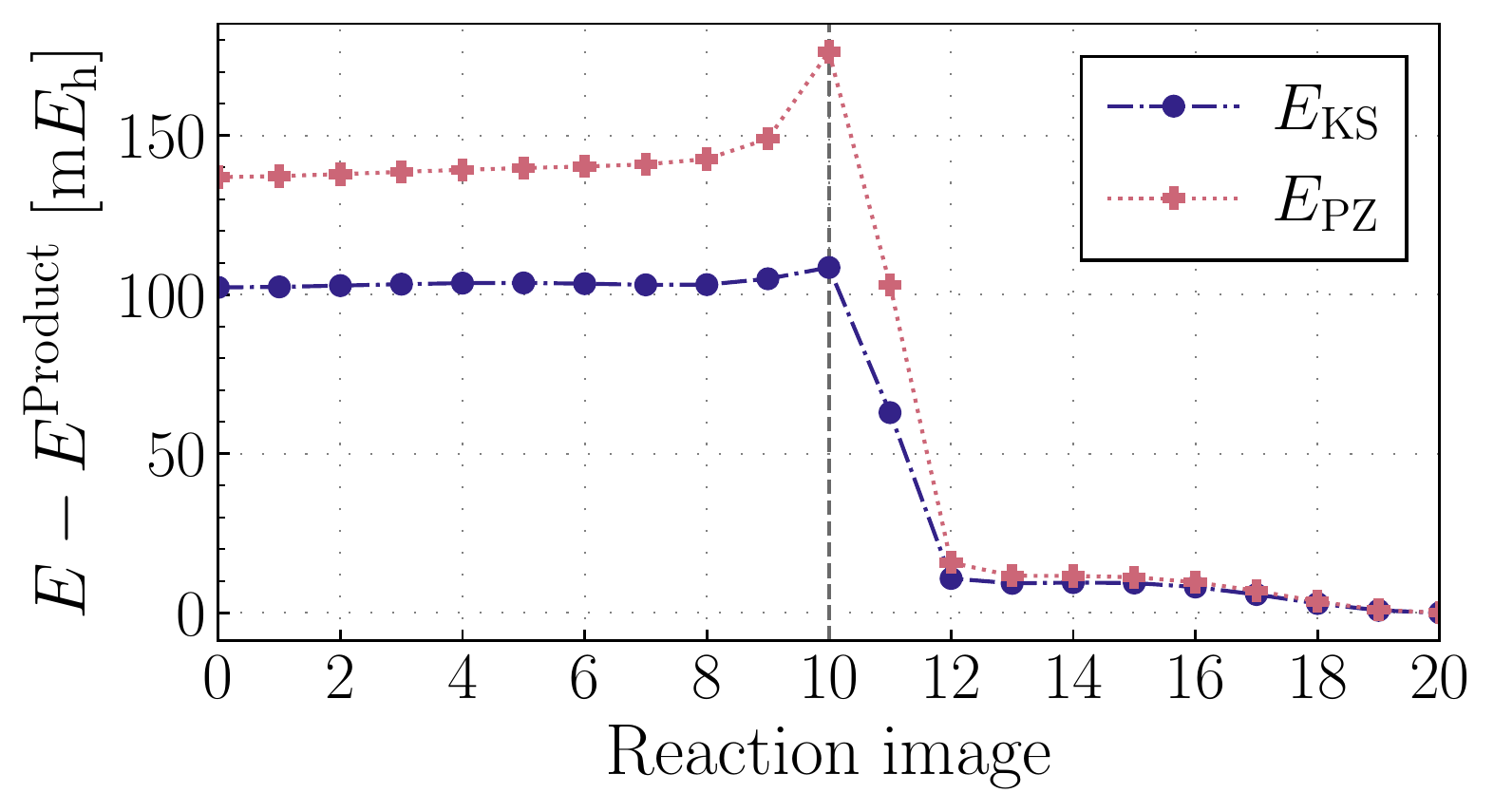}
		\caption{Total \ac{ks} energies $E_\mathrm{KS}$ and \ac{flosic} energies $E_\mathrm{PZ}$, shifted by the respective total energy of the product. Note that the \ac{flosic} energies are in any case lower than the \ac{ks} energies. See the supplementary material for the unshifted energies.}
		\label{fig:e_tot}
	\end{subfigure}

	\begin{subfigure}{\linewidth}
		\centering
		\includegraphics[width=\linewidth]{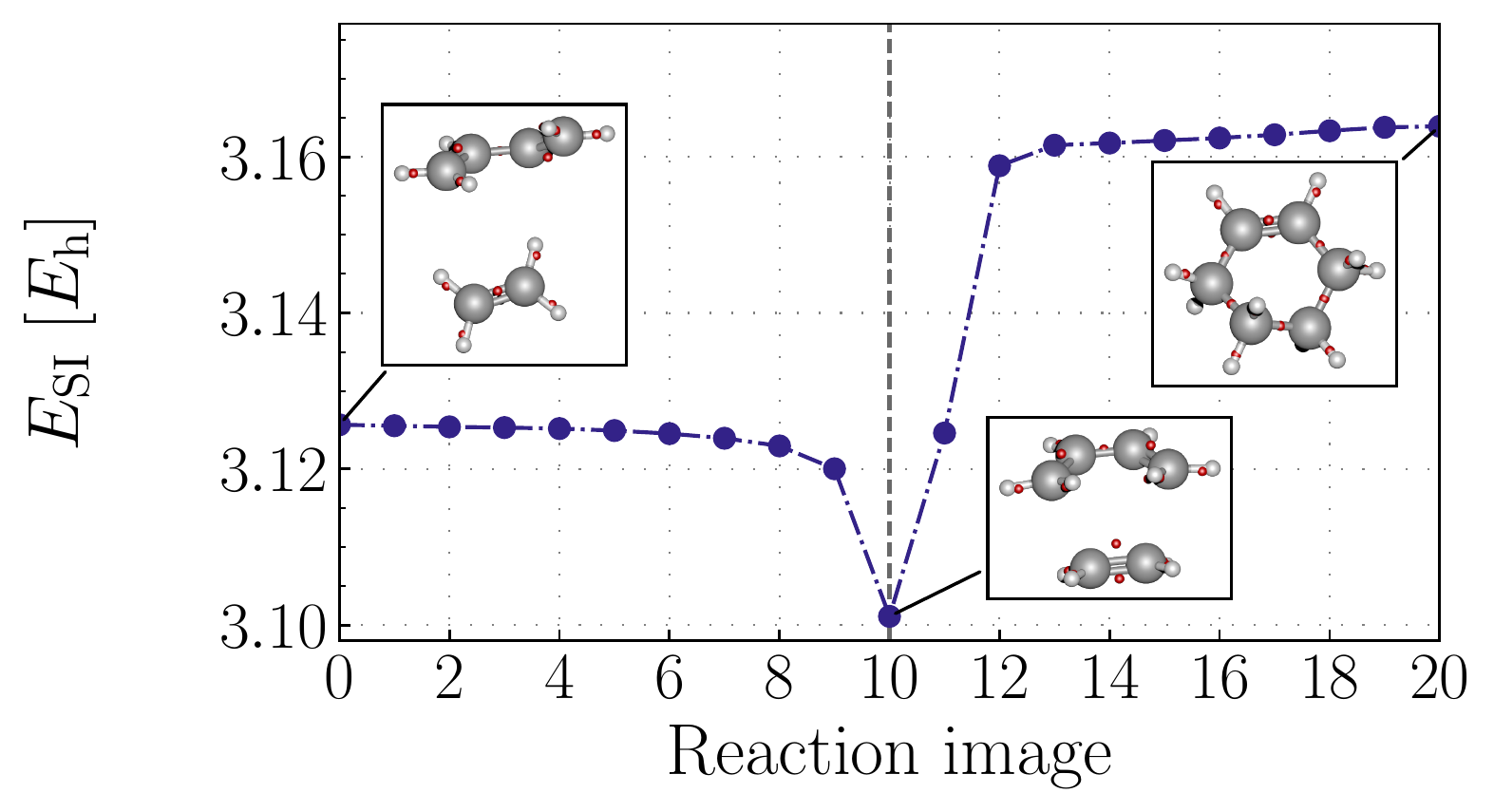}
		\caption{\ac{flosic} \acp{sie} $E_\mathrm{SI}$.}
		\label{fig:e_sie}
	\end{subfigure}

	\caption{Energies along the reaction images. The \ac{ts} is indicated at image 10 via a dashed line. The respective structures for the reactants, \ac{ts}, and product are sketched in (b) accordingly, with carbon atoms colored in grey, hydrogen in white, and the \acp{fod} in red. Only one spin
channel has been displayed since the \ac{fod} positions for these images coincide for both spin channels.}
	\label{fig:energetics}
\end{figure}

The total \ac{ks} energies $E_\mathrm{KS}$ for the sampled reaction are displayed in \figref{e_tot}, shifted by the total energy of the product. As expected, one can indicate the \ac{ts} as the \ac{hei}. Calculating the \ac{pz} \ac{sie} from the \ac{flosic} calculations (see \eqref{pzsic}), one obtains \figref{e_sie}. Firstly, one can see that the \ac{sie} is larger for the product than for the reactant. Interestingly, the \ac{sie} gets minimal for the \ac{ts}. However, the smaller \ac{sie} for the \ac{ts} is in line with the expectation from \refref{Shahi2019}. Applying this correction to the \ac{ks} energies, one obtains the \ac{flosic} energies $E_\mathrm{PZ}$ as seen in \figref{e_tot}. The reaction barrier increases for the corrected energies, agreeing with other research \cite{Patchkovskii2002, Johansson2008}.

\begin{figure}[hbp]
	\centering
	\includegraphics[width=\linewidth]{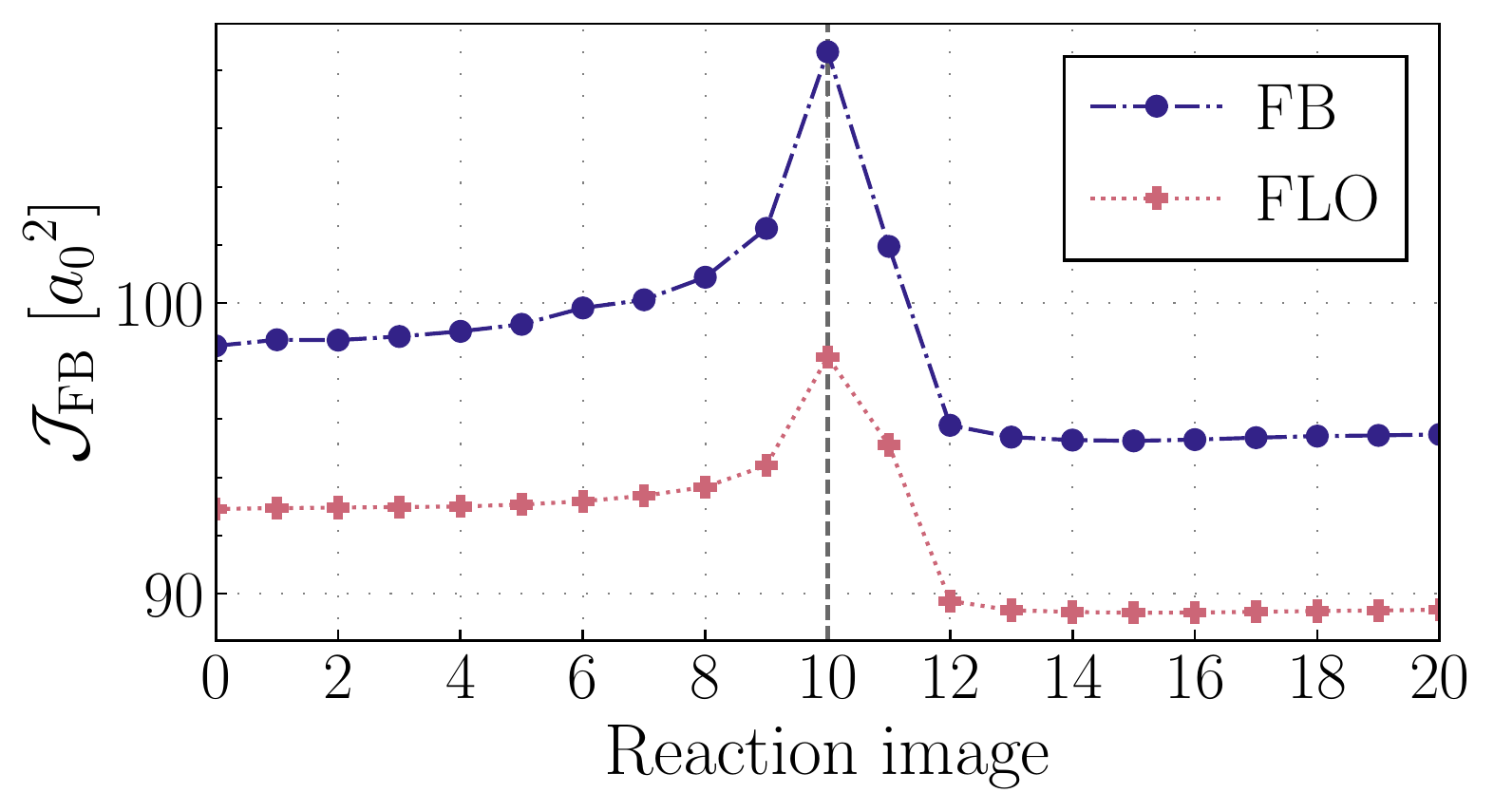}
	\caption{Orbital variances $\mathcal{J}_\mathrm{FB}$ of occupied \ac{fb} orbitals and \acp{flo} along the reaction images. The \ac{ts} is indicated at image 10 via a dashed line.}
	\label{fig:spread}
\end{figure}

\figref{spread} displays the orbital variances $\mathcal{J}_\mathrm{FB}$ (see \eqref{spread}) of the occupied \ac{fb} orbitals and \acp{flo}. The \acp{flo} are more localized than the \ac{fb} orbitals, but both show a similar trend. The figure roughly resembles the trend of the negative \ac{sie} in \figref{e_sie}. Accordingly, as the \ac{sie} gets minimal at the \ac{ts}, the \ac{ts} is also the most delocalized state as indicated by the maximal value of the \ac{fb} cost function (\figref{spread}). The bond breaking, as indicated previously by the \acp{abp} (see \figref{fod_educt}), can also be seen in the orbital variances of all localized orbitals which provide a measure of delocalization (see \eqref{spread}). The orbital variances of the remaining localized orbitals can be found in the supplementary material.

Similar behavior can be seen in other properties, like the absolute electric dipole moment $\mu$ for both the \ac{ks} and the \ac{flosic} densities, shown in \figref{dip}, or the ionization potential (see the supplementary material).
As proposed in \refref{Trepte2021}, the absolute value of the dipole moment is a simple descriptor of the density. We now find that this simple descriptor is also able to correctly monitor density changes leading to bond breaking and formation in this reaction. The dipole moment can indicate the re-arrangement of the density due to the relocation of the electrons, which is the actual bond breaking.

\begin{figure}[hbp]
	\centering
	\includegraphics[width=\linewidth]{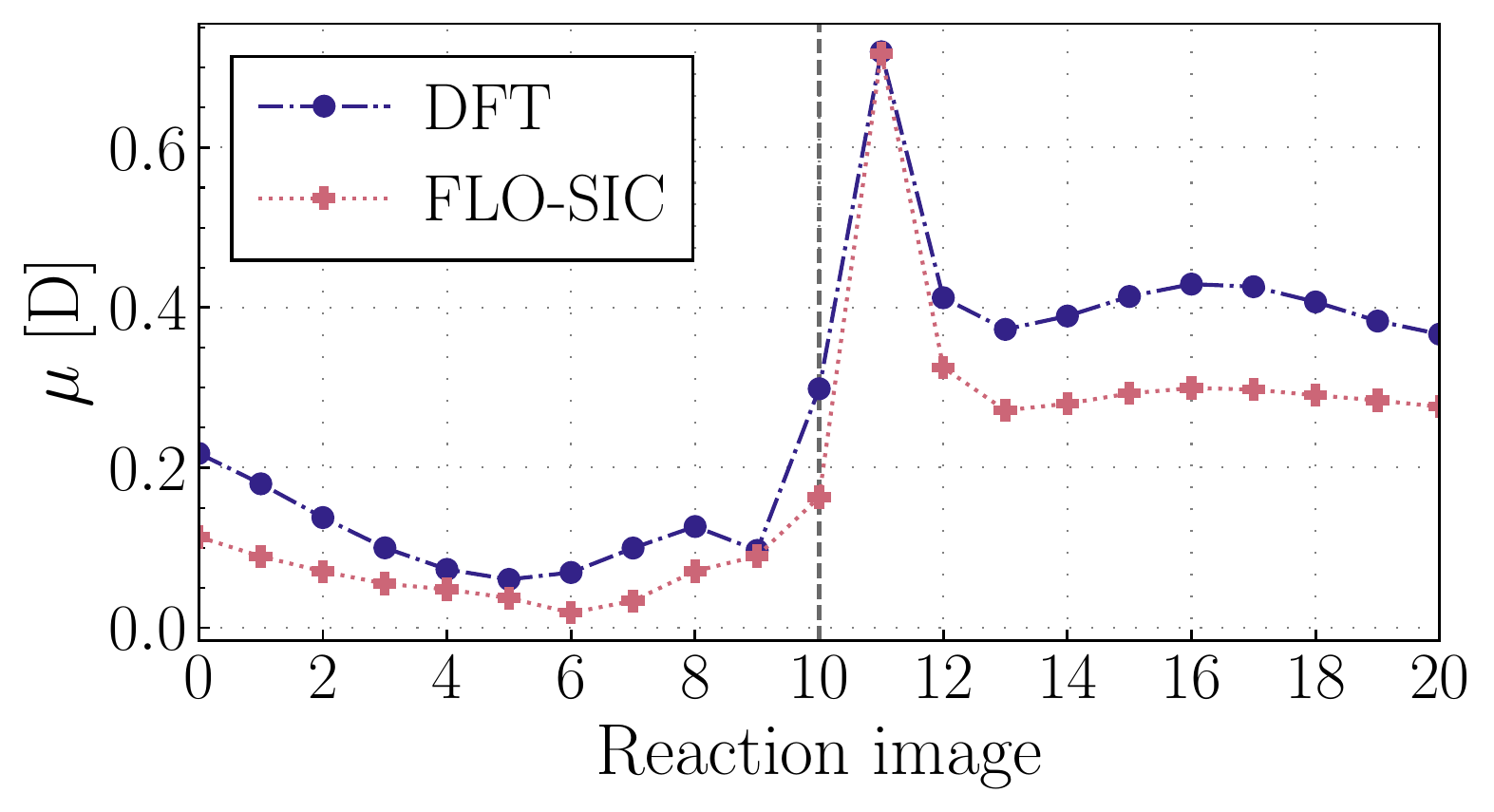}
	\caption{Absolute electric dipole moments $\mu$ of the systems along the reaction images, both for the \ac{dft} and the \ac{flosic} densities. The \ac{ts} is indicated at image 10 via a dashed line.}
	\label{fig:dip}
\end{figure}

\begin{figure}[htbp]
	\centering
	\includegraphics[width=\linewidth]{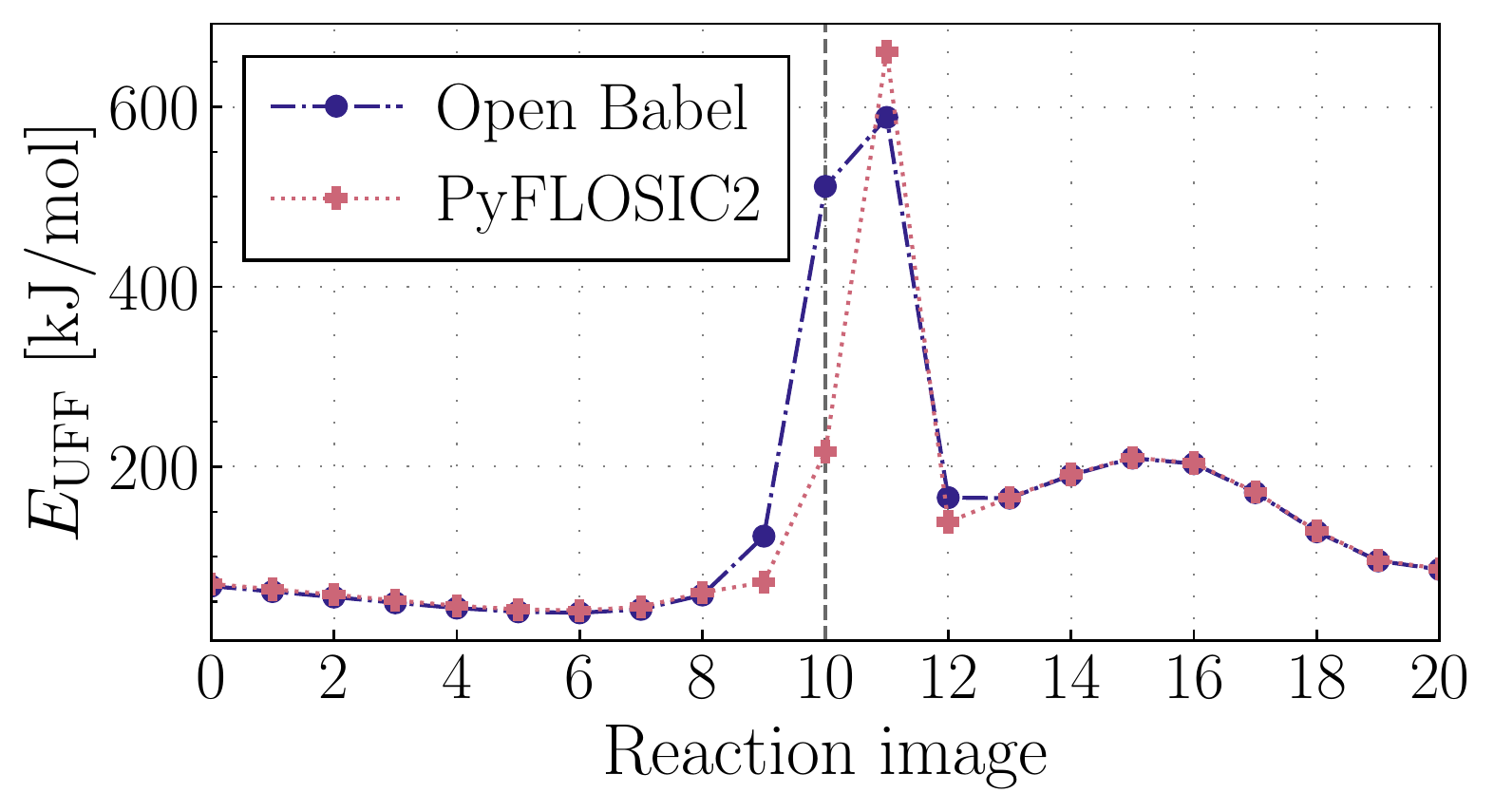}
	\caption{\Ac{uff} energies $E_\mathrm{UFF}$ of the systems along the reaction images using \pyflosic{} and \obabel{}. The \ac{ts} is indicated at image 10 via a dashed line.}
	\label{fig:e_uff}
\end{figure}

\vfill\eject
Additionally, we investigated the possibility to monitor bond changes using \ac{uff} energies \cite{Rappe1992}. In \figref{e_uff} one can see the calculated \ac{uff} energies $E_\mathrm{UFF}$ using \pyflosic{} and \obabel{} \cite{OBoyle2011}. \pyflosic{} utilizes a novel bond perception based on the \ac{fod} positions. Using nearest-neighbor relations between nuclei and \acp{fod}, each \ac{fod} gets classified as core, bond, or lone \ac{fod}. This information enables the calculation of the bond order matrix and the determination of the local chemical environment of each atom in a molecule. Since the centroids give the same bond order as the \acp{fod}, they will result in the same \ac{uff} energies.

Interestingly, the bond assessment for the \ac{uff} based on \acp{fod} delivers energies resembling the trend of the dipole moment $\mu$ (compare to \figref{dip}). While for the reactant and product state the \ac{abp}-based energies agree with the \obabel{} derived ones, there are significant differences around the \ac{ts}. The \ac{abp}-based bond perception describes the concerted change of the bond order better than the derived bonds from \obabel{}. Further investigations are necessary for a generalization of these findings. Despite this, the \ac{abp}-based \ac{uff} seems to be a promising candidate for an efficient scalar function tracking bond formation and breaking for similar systems.

\subsection{Verification of computational parameters}
To verify the qualitative trend of the found results, additional \ac{dft} calculations have been performed on the optimized structures, similar to the verification segment in \refref{Trepte2021}. To analyze the effect of the exchange-correlation functional, calculations have been performed utilizing the \acf{gga} functional PBEsol \cite{Perdew2008} and the \ac{mgga} functional r$^2$SCAN \cite{Furness2020}.
The main result is that independent of the functional, one can still see the bond formation in the dipole moment and the ionization potential. Moreover, the orbital variances of \ac{fb} orbitals display the same trends, while the $d_\mathrm{max}$ of \ac{fb} orbitals are mostly unaffected by the choice of the functional.

To analyze the effect of the basis set, \ac{dft} calculations using the family of polarization consistent basis sets pc-$n$ \cite{Jensen2001, Jensen2002, Jensen2002a} have been performed. Namely, the split-valence, double-$\zeta$, triple-$\zeta$, quadruple-$\zeta$, and quintuple-$\zeta$ basis sets pc-0, pc-1, pc-2, pc-3, and pc-4 have been investigated.
One finds a suitable agreement for pc-1 with the larger basis sets. Similar to the analysis of the functional, the $d_\mathrm{max}$ of \ac{fb} orbitals remain largely unaffected by the basis set choice. Accordingly, all basis sets recover the same trend for the dipole moment, ionization potential, and orbital variances.

Additionally, to confirm that there is no bias in the values of $d_\mathrm{max}$ due to the starting guess of the \ac{fod} optimization, an exemplary calculation using \ac{er} orbitals has been performed. As expected, both configurations result in the same optimized \ac{fod} configuration and therefore in the same values of $d_\mathrm{max}$, while the \ac{fb} guess needs fewer iterations to converge.

Finally, to ensure the reproducibility of our findings, reference \ac{dft} and \ac{flosic} calculations were performed with \chillijl{}.
The calculated \ac{dft} and \ac{flosic} energies along the reaction path show the same trend as obtained by \pyflosic{}. This accentuates that the presented \ac{flosic} trends are reproducible. The full analysis of all reference calculations can be found in the supplementary material.

\section{Summary \label{sec:summary}}
\acresetall
In this article, we investigated bond formation and breaking exemplary for the well-known Diels-Alder reaction from a self-interaction and local orbital perspective, using \ac{neb} calculations and optimizing the corresponding \ac{ts} using \ac{dft}. It has been shown that the \acp{fod} and centroids of different localized orbitals, i.e., \acp{flo}, \ac{fb}, and \ac{er}, can be used to describe and display the bonding situation not only in the reactant and product state but along the reaction as well. All \acp{abp} show a similar trend and indicate the bond formation and breaking at the same reaction image. More precisely, we find that the bond order changes exactly once along the reaction path while going from the \ac{ts} to image \tsp{}. This clearly shows that the concerted mechanism of the Diels-Alder reaction can be described using \acp{abp}. It has been shown that the (maximum) \acp{abp} distance from the middle point of the bond axis can be used to indicate the bond formation and breaking. We showed that for this reaction \ac{fb} orbitals are the best starting point from the tested centroids to generate initial \acp{fod}, utilizing the \pycom{} procedure.

\ac{dft} and \ac{flosic} calculations have been performed along the same reaction. In agreement with other research \cite{Patchkovskii2002, Johansson2008}, it has been shown that the reaction barrier gets raised when applying the \ac{sic} and the \ac{sie} gets minimal for the \ac{ts}. The bond formation can be observed in different properties like the absolute value of the dipole moment or the \ac{uff} energy derived from \acp{abp}.

For the investigated reaction, a Lewis configuration for the \acp{fod} was found. For any reaction having a Lewis configuration, one can expect similar behavior for the proposed measures of tracking bond changes. However, there exist cases where instead of Lewis configurations, Linnett double-quartet structures are preferred in \ac{flosic} \cite{Trepte2021, Liebing2022}. For these cases, the \acp{flo} and contained bonding information are vastly different from the localized orbitals of \ac{dft} calculations.
Thus, it could be interesting to apply the proposed measures to investigate reactions involving Linnett double-quartet structures.

The main finding of this work is that fully optimized \acp{fod} and centroids of localized orbitals are able to describe the bond formation and bond breaking for the analyzed Diels-Alder reaction. While other research has hinted at the bond information contained in \ac{fod} positions, the centroids are often more accessible. Such descriptors can be helpful to categorize the bonding situation in various applications, e.g., when analyzing self-healing materials. Therefore it is promising that they show similar trends. Whether these trends continue for more complex reactions should be investigated in future research.

\section*{Acknowledgments}
The authors thank Dr.\,Johannes Steinmetzer for helpful discussions and technical support regarding \pysis{}. The authors thank the Universitätsrechenzentrum of the Friedrich Schiller University Jena for computational time and support. W. T. Schulze and S. Gräfe highly acknowledge funding by the Deutsche Forschungsgemeinschaft (DFG, German Research Foundation) - Research unit FuncHeal, project ID 455748945 - FOR 5301 (project P5).
S. Schwalbe has been funded by the Deutsche Forschungsgemeinschaft (DFG, German Research Foundation) - Project ID 421663657 - KO 1924/9-2.
S. Schwalbe thanks Dr.\,Sebastian Borrmann for technical support and the HPC Freiberg for computational time.
The authors thank Dr.\,Susi Lehtola for comments on the original manuscript.
The authors thank an anonymous reviewer whose comments guided us to fix the \ac{fb} cost function in \pyscf{}.

\section*{Supplementary material}
Figures that display all \acp{abp} for selected structures can be found in the supplementary material. The summarized results of the verification calculations using \chillijl{}, and the calculations using different functionals and basis sets are discussed. Additionally, figures for the values of $d_\mathrm{max}$ for all \acp{abp}, the ionization potentials, and \ac{uff} contributions from \pyflosic{} and \obabel{} are listed. A reference implementation to calculate the isovalues from \figref{geometry} is included. More information about the symmetry of the optimized nuclear geometries, the localization procedure, bond perception, and the definition of the electric dipole moment are appended.

\section*{Author declarations}
\subsection*{Conflict of Interest}
The authors have no conflicts to disclose.

\section*{Data availability}
The data that support the findings of this study are available within the article and its supplementary material. The initial and optimized structures, \ac{neb} images, and images with \acp{abp} included can be found openly available under \url{https://gitlab.com/wangenau/bond_formation_supplementary}.

\bibliography{refs.bib}
\end{document}